\documentclass[aps,prl,preprint,groupedaddress]{revtex4}

\usepackage{graphics}
\usepackage{dcolumn}
\usepackage{bm}

\begin{document}


\title{ Revisiting the Perfect Lens}


\author{Weiguo Yang}
\email[]{wyang@email.wcu.edu}
\affiliation{ Department of Engineering \& Technology, Western Carolina University, Cullowhee, NC 28723, USA}

\author{ John O. Schenk and Michael A. Fiddy}
\email[]{mafiddy@uncc.edu}
\affiliation{Center for Optoelectronics and Optical Communications, University of North Carolina at Charlotte, NC 28223, USA}


\date{\today}

\begin{abstract}
We carefully examine the negative refractive index slab perfect lens theory by Pendry\cite{pendry2000} and point out an inconsistency that can be resolved. As a result, we find negative index slabs do not amplify or enhance evanescent waves and therefore they do not make a perfect lens in the sense that was originally suggested. 
\end{abstract}


\maketitle


\section{Introduction}


In his seminal paper \cite{pendry2000}, Pendry derived the fascinating result that a slab of negative index material, despite the fact that evanescent waves would decay away from the interfaces, amplifies the evanescent waves sustaining them through the material to the other side.   Since these evanescent waves carry high spatial frequency information about an object, this phenomenon opens up the opportunity of realizing a higher resolution lens, perfect in the ideal case. Initially, this counter-intuitive result was met with some controversy \cite{ctr2,ctr3,ctr4,ctr5,ctr6,ctr7,ctr8,ctr9}.  Later, experimental results indicated support for the amplification of evanescent waves in negative index slab, notably by Liu \cite{Liu6}.  This quieted the controversies and the negative index perfect lens concept was on a solid enough foundation that research efforts for practical applications are being pursued all over the world \cite{Science7}.   However, there is a question arising in Pendry's paper concerning the expressions derived for the reflection and transmission coefficients at the negative index boundary.  In this paper, we review the derivation of the result given by Pendry and point out the expressions in question.  We then provide alternative and more consistent solutions to the problem which show that the negative index slab of n= -1 ($\epsilon$ = -1 and  $\mu$= -1) does not make a perfect lens. We will also discuss the experimental results of Liu\cite{Liu6} and argue that the experimental results did not provide an unequivocal proof of the "amplification" or "enhancement" of the evanescent waves by a negative index slab. 

\section{Pendry's Negative Index Perfect Lens Formulism}

\begin{figure}
\centerline{\resizebox{3.5in}{!}{\includegraphics{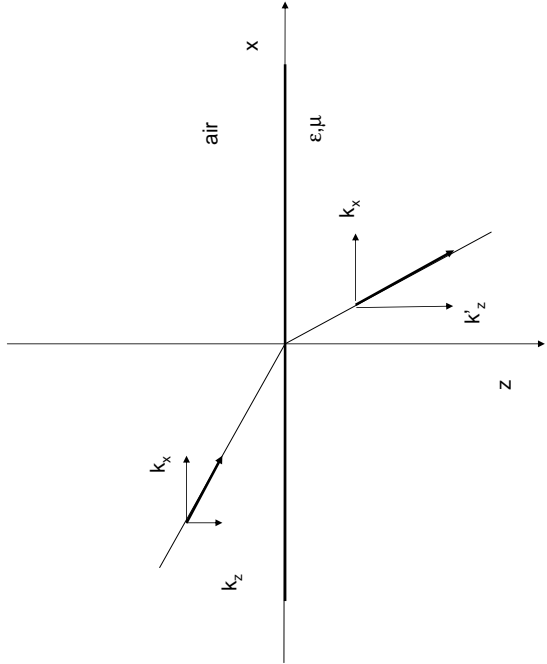}}}
 \caption{\label{1}Refraction at the interface of two different materials. }
 \end{figure}

Figure\ref{1} shows an interface between two different materials. The $(x-z)$ plane is chosen to be the plane that contains the propagation vector $\mathbf{k}=(k_x,k_z)$ and a normal of the interface. Following Pendry's notation \cite{pendry2000}, when $k_x<|\mathbf{k}|$, $k_z$ is a real number and plane waves $e^{i\mathbf{k}\cdot\mathbf{r}-i\omega t}$ are propagating waves. When $k_x>|\mathbf{k}|$, $k_z$ is an imaginary number and plane waves $e^{i\mathbf{k}\cdot\mathbf{r}-i\omega t}$ have an imaginary wavevector component and become evanescent waves. We follow the Ref.\cite{pendry2000}'s derivation (from Eq.(13) to Eq.(18) in Ref.\cite{pendry2000}) and point out an inconsistency.

The derivation assumes S-polarized light in vacuum, whose electric field is given by   
\begin{equation}
E_{0S+}=[0,1,0]\exp(ik_z z+ik_x x-i\omega t)
\label{stq}
\end{equation}
where the wavevector $k_z=+i\sqrt{k_x^2-\omega^2c^{-2}}$, $\omega^2c^{-2}<k_x^2$ is an evanescent wave with exponential decay along positive $z$ direction. The electric field of the reflected light, indicated by the $-$ subscript, is given by,
\begin{equation}
E_{0S-}=r[0,1,0]\exp(-ik_z z+ik_x x-i\omega t)
\end{equation}
where $r$ is the reflection coefficient and the transmitted electrical field is given by,
\begin{equation}
E_{1S+}=t[0,1,0]\exp(ik^\prime_z z+ik_x x-i\omega t)
\end{equation}
where 
\begin{equation}
k^\prime_z=+i\sqrt{k_x^2-\epsilon\mu\omega^2c^{-2}},\; \epsilon\mu\omega^2c^{-2}<k_x^2.
\label{kprime}
\end{equation}

As stated in Pendry's original paper as well as in several other related paper \cite{pendry2000, ctr3}, causality requires that $k^\prime_z$ is chosen this form because the evanescent wave must decay away from the interface. 

The boundary conditions implied by the transmission and reflection coefficients (Eq. 18 in Ref.\cite{pendry2000}) are the usual continuity conditions, \begin{eqnarray}
(E_{\parallel})_1=(E_{\parallel})_2 & (\epsilon E_{\perp})_1=(\epsilon E_{\perp})_2,\nonumber \\
(H_{\parallel})_1=(H_{\parallel})_2, &(\mu H_{\perp})_1=(\mu H_{\perp})_2,
\label{bndc}
\end{eqnarray}


The magnetic field $H$ and the electric field $E$ are linked by Maxwell's equations and given quasimonochromatic waves one has $k\times E=\omega\mu\mu_0 H$. It follows that $\omega\mu_0(H_{0S+})_x=-k_zE_{0S+}$, and $\omega\mu_0(H_{0S+})_z=k_xE_{0S+}$. For reflected field, $\omega\mu_0(H_{0S-})_x=k_zE_{0S-}$, and $\omega\mu_0(H_{0S-})_z=k_xE_{0S-}$, and for the transmitted field, $\omega\mu\mu_0(H_{1S+})_x=-k_zE_{1S+}$, and $\omega\mu\mu_0(H_{1S+})_z=k_xE_{1S+}$.  Combining these relations with the boundary conditions, one has equations for the transmitted and reflected fields,


\begin{eqnarray}
E_{0S+}+E_{0S-}&=&E_{1S+} \nonumber \\
-\mu k_zE_{0S+}+\mu k_zE_{0S-} &=&- k^\prime_zE_{1S+},
\label{eq21}
\end{eqnarray}

Solving these equations gives,

\begin{eqnarray}
\frac{E_{1S+}}{ E_{0S+}}&=t=&\frac{2\mu k_z}{\mu k_z+ k^\prime_z} \nonumber \\
\frac{E_{0S-}}{ E_{0S+}}&=r=&\frac{\mu k_z- k^\prime_z }{\mu k_z+ k^\prime_z} 
\label{edq}
\end{eqnarray}
which are Eq.(18) in Ref.\cite{pendry2000}.

The sequence of steps from Eq.\ref{stq} to Eq.\ref{edq} (or Eq.(13) to Eq.(18) in Ref.\cite{pendry2000}) seems flawless. However, if one takes $k_z$ and $ k^\prime_z$ as defined in previous text, for the case where $\epsilon=-1$ and $\mu=-1$, $k^\prime_z=k_z$ and $\mu k_z+ k^\prime_z=0$.  In Pendry's paper, the transmission is calculated through the front and back surfaces of a slab of thickness $d$, with the multiple scattering included, and the limit taken as $\epsilon$ and $\mu$ tend to -1. This yields the result that $\exp(-i k^\prime_z d)=exp(-ik_z d)$ and hence that evanescent and propagating waves contribute to the resolution of the image. The zero denominator in Eq.\ref{stq}, however, does warrant a closer look to determine its significance. The origin of this singularity can be traced back to the defining equations Eq.\ref{eq21}, where for the special case of $\epsilon=-1$ and $\mu=-1$ one has,
\begin{eqnarray}
E_{0S+}+E_{0S-}&=&E_{1S+} \nonumber \\
E_{0S+}- k_zE_{0S-} &=&- E_{1S+},
\end{eqnarray}

This set of equations is self-contradictory when the incident field $E_{0S+}\neq 0$.  Since Eq.\ref{eq21} are derived from boundary conditions, this paradox is a result that the form for $ k^\prime_z$ (Eq.\ref{kprime}) is not compatible with the boundary conditions Eq.\ref{bndc}. 

\section{A simple solution to the apparent paradox}

Consider the situation for which the medium has a negative refractive index for doubly negative materials ($\epsilon <0$ and $\mu<0$), and let us assume, for S-polarization, the evanescent wave in this medium takes the form,
\begin{equation}
E_{1S+}=t[0,1,0]\exp(ik^\prime_z z+ik_x x+i\omega t+i\phi)
\end{equation}
where $k^\prime_z$ is still given by Eq.\ref{kprime} and $\phi$ is a constant phase.  The evanescent wave should decay away from the inferface, which is the same causality requirement stated before and used by Pendry \cite{pendry2000,ctr3}. Note that both $\exp i(\mathbf{k}\cdot\mathbf{r}-\omega t)$ and $\exp i(\mathbf{k}\cdot\mathbf{r}+\omega t)$ are solutions to the Helmholtz equation $\nabla^2E+\epsilon\mu\omega^2c^{-2}E=0$. The difference between these is that their phase velocities are in opposite directions. Therefore, if we choose one form for positive index of refraction medium, we must choose the other form for negative index of refraction medium. 

Now since the boundary conditions Eq.\ref{bndc} apply to the real time dependent fields, we can express all fields in their real time dependent values,
\begin{equation}
E_{0S+}=[0,1,0]\exp(-|k_z|z)\cos(k_x x-\omega t+\phi_0),
\end{equation}
where $\phi_0$ is a phase constant. The $x$ component of the magnetic field is accordingly,
\begin{equation}
\omega\mu_0 (H_{0S+})_x=-|k_z| \exp(-|k_z|z)\cos(k_x x-\omega t+\phi_0+\pi/2),
\end{equation}
The $\pi/2$ phase difference between the $y$ component of the electric field and the $x$ component of the magnetic field is characteristic of evanescent waves since they do not carry energy flow in the $z$ direction.  
The $z$ component of the magnetic field is in phase with the $y$ component of the electric field,
\begin{equation}
\omega\mu_0 (H_{0S+})_z=k_x \exp(-|k_z|z)\cos(k_x x-\omega t+\phi_0),
\end{equation}
which indicates that there will be energy flow in the $x$ direction if $k_x$ is nonzero. 

Similarly for the reflected wave, we have,
\begin{equation}
E_{0S-}=r[0,1,0]\exp(|k_z|z)\cos(k_x x-\omega t+\phi_{0-}),
\end{equation}
where $\phi_{0-}$ takes into account a possible different phase constant and,
\begin{eqnarray}
\omega\mu_0 (H_{0S-})_x&=&|k_z| r\exp(|k_z|z)\cos(k_x x-\omega t+\phi_{0-}+\pi/2),\\
\omega\mu_0 (H_{0S-})_z&=&k_x r\exp(|k_z|z)\cos(k_x x-\omega t+\phi_{0-}),
\end{eqnarray}

The electric field of the refracted wave in the negative refraction index medium in its time dependent form is
\begin{equation}
E_{1S+}=t[0,1,0]\exp(-|k^\prime_z|z)\cos(k_x x+\omega t+\phi),
\end{equation}
and the components of the magnetic field are,
\begin{eqnarray}
\omega\mu\mu_0 (H_{1S+})_x&=&-|k^\prime_z| t\exp(-|k^\prime_z|z)\cos(k_x x+\omega t+\phi+\pi/2),\\
\omega\mu\mu_0 (H_{1S+})_z&=&k_x t\exp(-|k^\prime_z|z)\cos(k_x x+\omega t+\phi),
\end{eqnarray}

The conditions in Eq.\ref{bndc} require that at the boundary,
\begin{eqnarray}
\cos(-\omega t+\phi_0)+r\cos(-\omega t+\phi_{0-})&=&t\cos(\omega t+\phi),\nonumber \\
-|k_z|\cos(-\omega t+\phi_0+\pi/2)+ |k_z|r\cos(-\omega t+\phi_0+\pi/2)&=&-\frac{|k^\prime_z}{\mu}t\cos(\omega t+\phi+\pi/2),
\label{ndf}
\end{eqnarray}

The $x$ component of the k-vector still has to be the same for all three fields in order for the boundary condition to support non-zero solutions.  Steady state solutions require that transmission coefficient $t$ and reflection coefficient $r$ do not depend on time. Such solutions are possible only if $\phi_{0-}=\phi_0$ and $\phi=-\phi_0$ since the phase velocities have opposite signs across the boundaries and the phase behavior reflects this. Consequently, Eq.\ref{ndf} become,
\begin{eqnarray}
1+r&=&t \nonumber \\
-|k_z|+|k_z|r&=&\frac{|k^\prime_z|}{\mu}t
\end{eqnarray}
which give the transmission and reflection coefficients for evanescent waves at the interface with the negative index medium,
\begin{eqnarray}
t&=&\frac{2\mu|k_z|}{\mu|k_z|-|k^\prime_z|} \nonumber \\
r&=&\frac{\mu|k_z|+|k^\prime_z|}{\mu|k_z|-|k^\prime_z|},
\end{eqnarray}

This solution does not have the singularity as Pendry's result in Eq.\ref{edq} for the special case of $\epsilon=-1$ and $\mu=-1$. Consider the implications of this form for the reflection and transmission coefficients for evanescent waves at an interface between positive and negative index when $\epsilon=-1$ and $\mu=-1$, we find that $t=1$ and $r=0$, and following the rest of Pendry's derivation, the transmission through both surfaces of a negative index slab is $T_s=\exp(ik^\prime_z d)$, which is now an attenuation rather than amplification or enhancement for the evanescent wave.  We have to conclude from this that a negative index slab having $\epsilon=-1$ and $\mu=-1$ does not amplify or enhance the evanescent waves. In fact, the evanescent wave simply decays through such a slab.

A similar result can be obtained for P polarization, where we have,
\begin{eqnarray}
t_p&=&\frac{2\epsilon|k_z|}{\epsilon|k_z|-|k^\prime_z|}\nonumber \\
r_p&=&\frac{\epsilon|k_z|+|k^\prime_z|}{\epsilon|k_z|-|k^\prime_z|},
\end{eqnarray}
and similarly, there is no amplification or enhancement of the P polarized evanescent waves when $\epsilon=-1$ and $\mu=-1$. Again one has $T_p=\exp(ik^\prime_z d)$ indicating an attenuation.

\section{Another equivalent interpretation}
Consider equivalent solutions to the Helmholz equation, $\exp[-i(\omega t-\mathbf{k}\cdot\mathbf{r})]$ and $\exp[-i(\omega t+\mathbf{k}\cdot\mathbf{r})]$, the difference between the two is that their phase velocities are in opposite directions.  Using these solutions for positive and negative index media respectively, and following a similar approach to that presented in the previous section, there will be no apparent phase jumps, i.e., $\Delta\phi=\phi-\phi_0=0$.  The boundary conditions Eq.\ref{bndc} are now applicable to complex fields as well and the solutions for the transmission and reflection coefficients take the usual form,
\begin{eqnarray}
t&=&\frac{2\mu k_z}{\mu k_z+k^\prime_z}\nonumber \\
r&=&\frac{\mu k_z - k^\prime_z}{\mu k_z+k^\prime_z}  
\end{eqnarray}
for S polarization and
\begin{eqnarray}
t_p&=&\frac{2\epsilon k_z}{\epsilon k_z+k^\prime_z}\nonumber \\
r_p&=&\frac{\epsilon k_z - k^\prime_z}{\epsilon k_z+k^\prime_z},
\end{eqnarray}
for P polarization.
These are applicable to both propagating and evanescent waves.  In this case, however, the evanescent wave in the negative index medium that decays away from the interface must have the $z$ component defined by,
\begin{equation}
k^\prime_z=-i\sqrt{k_x^2-\epsilon\mu\omega^2c^{-2}},\; \;\; \epsilon\mu\omega^2c^{-2}<k_x^2,
\end{equation}

Consequently, for the special case where $\epsilon=-1$ and $\mu=-1$, $k^\prime_z=-k_z$ and we have $t=1$ and $r=0$ for both S and P polarizations.  Again the negative index slab does not amplify or enhance the evanescent waves. 

\section{Discussion}
As mentioned in the introduction, Ref.\cite{Liu6} is cited as an experimental verification of the amplification or enhancement of the evanescent wave in a negative index slab \cite{Science7}. The maximum transmission of a sliver slab, which serves as a lossy negative index slab, is plotted against the thickness of the slab in Fig.4 of Ref.\cite{Liu6}.  The apparent rapid growth for thicknesses less than 60 nm is attributed to the amplification of the evanescent waves and the decrease of the transmission for thicknesses larger than 60 nm is attributed to loss in the material.  This explanation can be interpreted differently.  If the evanescent wave is amplified by the slab by $\exp{-ik^\prime_z d}$  as suggested by Pendry \cite{pendry2000}, then taking into account the material loss $\alpha$ and assuming that the material loss is independent of the thickness of the material to the first order, then the total transmission for the slab would be $\exp(-ik^\prime_z d)\exp(-\alpha d)=\exp[(-ik^\prime_z-\alpha)d]$. Consequently we have two possibilities. One is that $(-ik^\prime_z-\alpha)>0$  and the 'gain' is larger than the loss and the transmission should keep increasing with the thickness $d$.  The alternative is that $(-ik^\prime_z-\alpha)>0$   and the 'gain' is smaller than the loss and the transmission should monotonically decrease with the thickness $d$.  The experimental results shown in Fig.4 of Ref\cite{Liu6} do not really conform to either case and we would humbly suggest do not really serve as a verification of the 'amplification' or enhancement of the evanescent waves in this negative index  slab. 
\bibliography{na}

\end{document}